# Direct Numerical Simulation for Parametric Vertical Vibration and Atomization of Sessile Drops

Imperial College London, Department of Chemical Engineering, South Kensington campus, London, SW7 2AZ

Mr. Debashis Panda (CID: 02062148)

MSc. Advanced Chemical Engineering with specialization in Process Systems Engineering

Supervisor: Prof. Omar K Matar, Head of Department of Chemical Engineering

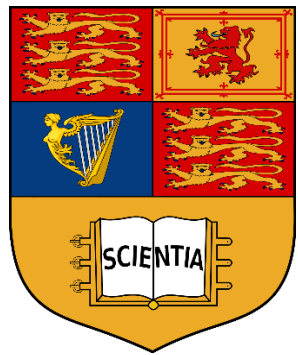

# Direct Numerical Simulation for Parametric Vertical Vibration and Atomization of Sessile Drops

Debashis Panda, Lyes Kahouadji, Omar K Matar*

*Department of Chemical Engineering, Imperial College London, South Kensington Campus, SW7 2AZ, United Kingdom*

ABSTRACT

Sessile drop vertical vibrations and atomization is commonly analyzed by simpler unconstrained instabilities like Faraday waves and hence, detailed studies of the phenomenon are still an open question to the fluid dynamics community. We address two critical gaps using direct numerical simulation, i.e., (i) the cause of subharmonic response of azimuthal waves, and (ii) the dependence of vertical vibrations to the universal pinch-off regime in the atomization. First, a high mode excitation of a large drop of 100 $\mu L$ is implemented. It is found that the subharmonic response is not via Faraday waves, rather it is the harmonics of the interfacial waves that induce azimuthal waves near the contact line. An analogous self-inducing dynamical system is formalized between the tip and near the contact line to elucidate the mechanism. Second, a low mode excitation of small drop of 30 $\mu L$ is used to critically explain the cause of primary pinch-off via crater formation and secondary pinch-off via ligament retraction. The pinch-off test in both cases revealed that the primary pinch-off is in visco-capillary scaling regime, while the secondary pinch-off is affected by the inertial forces via oscillatory body forces of vibration.

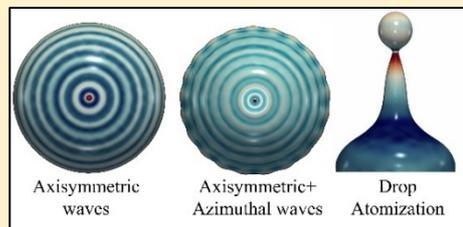

Key Highlights:

- First ever numerical simulation for high mode excitation of sessile drop at large amplitude.
- Subharmonic responses in azimuthal waves are not Faraday waves. It is rather formed via the harmonics of the interfacial waves- a self-induced dynamical system.
- Primary pinch-off follows visco-capillary universal scaling while secondary pinch-off is affected by the inertial oscillatory forces and forms satellite drops.

## 1. INTRODUCTION

Hydrodynamic instabilities such as Rayleigh oscillations on a spherical drop [1], [2], Rayleigh-Plateau breakup of cylinder [3]–[5], Faraday waves on a planar interface [6]–[8] are some of the classical problems encountered in the fluid dynamics community. These problems are emphasized as '*unconstrained*' instabilities as the fluid interfaces are not bounded by any solid surface. On the contrary, a growing interest in '*constrained*' instabilities (where uncompensated Young's stresses due to three-phase contact line bounds the fluid domain) has been observed from the past decade. The growing demand is a result of technological innovations in adaptive liquid lens [9], fuel injection [10], water electrolyzers [11] and drop atomization [12]–[15]. Drop atomization is commonly utilized in drug delivery and biomedical applications [16]–[18]. Further, fluid dynamics community is exploring numerous applications of drop atomization in biological applications such as modification of biomaterials in drug distributing transporters, oral vaccine delivery, and implant coating [19]. Biomimicry is also a new field of drop atomization application where tiny burgeoning robot flyers are mimicking mosquitoes for fast evaporation of moisture [20]. The mosquitoes employ a modified beat closest to the minimum harmonic mode, called the flutter strokes causing the atomization of moisture from the surface of the wings. Such flutter strokes are of amplitude 1 mm but at a frequency of 1kHz generating an acceleration of at least 250$g$ for the atomization of the drops in a few milliseconds.

Sessile drop vibration and atomization is a long-standing problem for the past 30 years. The first experimental study on the forced vibrational modes on such constrained capillary interfacial wave is conducted by Rodot et al. [21]. They investigated the oscillations of the drop levitated on a cylindrical substrate by following a method proposed by Plateau [5]. Unlike unconstrained drop vibration of radius $R$, where the first-mode resonant frequency is proportional to $R^{-1.5}$, they predicted that it is proportional to $R^{-2}$ in the constrained capillary drop. Noblin et al. [22], [23] are among the first few researchers who emphasized the importance of three phase contact-line. They studied vibration-induced sessile drop dynamics on polystyrene where they observed that at lower amplitude the drops exhibit axisymmetric, concentric, harmonic standing waves with pinned contact line (exhibiting contact-line hysteresis). Moreover, they observed non-axisymmetric modes with stick-slip contact line motion at higher amplitudes. On the contrary, Vukasinovic et al. [13], [24], observed pinned contact line even at higher amplitudes of high mode excitation. In their work, a large sessile drop is vibrated at 1kHz, and the amplitude is ramped in such a way that various modes of excitation are observed on the interface. Such discrepancies lead to an open discussion on the physics of the sessile drop vibrations.

A typical observation drawn from the experiments of Vukasinovic et al. [24] is the formation of azimuthal waves along the contact line beyond a threshold acceleration. They observed that unlike the axisymmetric harmonic standing waves, the azimuthal waves exhibited subharmonic response (i.e., half-frequency of parametric vibrations). Parametric interfacial waves with subharmonic response are a classical signature of Faraday waves. They related the finite depth linear stability problem for the Faraday waves to understand the subharmonic response. However, the critical acceleration for the azimuthal wave formation was higher than predicted with the analytical solution. They explored various possibilities to explain such disagreement e.g., capillary damping by contact line, interfacial damping by interfacial boundary layer and the viscous boundary layer. Upon proving that the capillary damping and interfacial damping are not the cause of the disagreement, they concluded that the viscous damping along the boundary layer in contact is the reason of higher threshold acceleration.

However, the above analysis of azimuthal waves as a classical signature of Faraday waves is misleading as Kumar [8] observed a crossover harmonic response when a fluid of depth lower than the viscous boundary layer is under parametric excitation. Hazra et al. [25] observed the presence of coupled harmonic and subharmonic response in case of modulated Rayleigh-Benard magnetoconvection. Moreover, they investigated that the threshold frequency response can be either purely harmonic or subharmonic depending on the Rayleigh number. Muller et al. [26] also observed that the thin fluid layers showcase harmonic response at its threshold over a wide range of frequency.

These discrepancies with the Vukasinovic et al. [24] can be explained with the following arguments: (a) There is a restriction on using the linear stability theory for Faraday waves which states that, "*the difference between the parametric frequency and the resonant frequency of the fluid system should be much lesser than the excitation frequency.*" It means that the excitation must be applied very near to the resonant frequency which is overlooked in various works on Faraday waves literature [27], [28]. In their work, the deviation from resonant frequency is accounted to be 5% which may explain the disagreement with the Faraday waves. (b) The sessile drop forms a spherical cap on the vibrating solid along with a three-phase contact line. Generalized Faraday waves is accounted when the interface is under perpendicular time-varying periodic acceleration as shown in **Fig. 1**. The time-varying periodic acceleration perpendicular to $xy$ plane for planar interface (see **Fig. 1(a)**) is the classical example of Faraday waves used in the community. Further, in case of curvilinear coordinates, the analogue of planar Faraday waves is showcased for sphere (see **Fig. 1(b)**). Recently, Ali-Hugo and Tuckerman [29], [30] explored the rich dynamics on the spherical analogue where the time varying periodic acceleration is in radial direction. They found axisymmetric, tetrahedral, cubic, $D_4$, icosahedral patterns on the sphere as the mode of vibrations is increased.

Another interesting analogue of the Faraday waves is on a cylindrical surface where the time-varying oscillation is in radial direction (see **Fig. 1(c)**). Patankar et al. [3], utilized such an analogue as a stabilizing strategy in the case of Rayleigh-Plateau instability on a cylindrical jet. On the contrary, in the case of a sessile drop, the vertical time-varying acceleration can lead to two (or three in case of three-dimensional) components, i.e., one perpendicular to the interface (radial direction) leading to Faraday waves, and the other horizontal time-varying acceleration providing oscillatory stresses on the liquid-gas interface. Hence, the mechanism of azimuthal wave formation along the contact line is not a cause of Faraday waves generated due to the vibrating surface (**Fig. 1(d)**). These qualitative reasonings provide evidence that a crucial gap is present in elucidating the mechanism of azimuthal waves on a sessile drop.

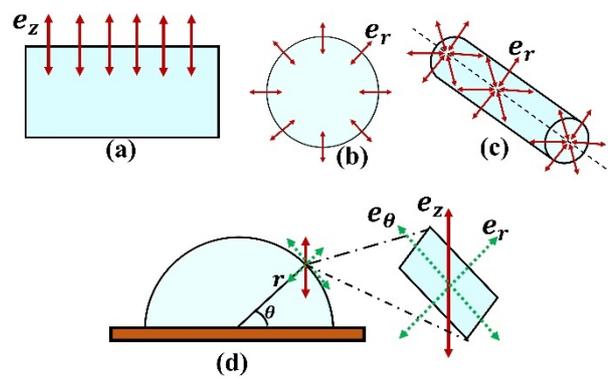

Fig. 1. Pictorial representation of Generalized Faraday waves for (a) Planar (b) Sphere (c) Cylinder (Note: the coordinates relevant to showcase Faraday waves are shown here) and (d) Vertical excitation of sessile drop decomposed into Faraday waves and oscillatory horizontal acceleration

Sessile drop atomization via vibration is therefore an open problem as not only the interfacial dynamics is complicated as explained above but also the non-linear effects presented by the viscous, surface tension and inertial forces govern the drop's necking and pinch-off dynamics. A vast literature concludes that the inertio-capillary forces usually govern the pinch-off mechanism in constrained as well as unconstrained drops, bubbles and soap films [31], [32]. The *spreading-like* law defines that the inertio-capillary regime follows 2/3 scaling law for drops and bubbles [33]. However, deviation from the scaling law is observed in the case of a bubble surrounded by water where it follows the 1/2 scaling law [34]. Even theoretical predictions concluded that such pinch-off dynamics occur in the Rayleigh regime with a scaling law of 0.56 [35]. However, such pinch-off events are experienced by a drop pulled away either from another drop or from a hydrophilic surface. In either of the cases, the drops are not forced by an external agent for example, the vibration induced via a surface. In our case, the constrained sessile drop is forced at a constant acceleration amplitude and frequency and the lag of reaction time to the drop's interface due to it inertia plays a vital role in the formation of conical spike which results in the pinch-off event. Such conical spike is similar to the works of Stone et al. [36], [37], where they showcased the breakup of extended liquid threads during relaxation. They concluded that the pinch-off dynamics in this case follow the visco-capillary regime and the inertial forces do not play a vital role. Moreover, it is extensively concluded that such scaling laws fail to fit under forced body forces. For example, in the case of drop coalescence the scaling law follows that Tanner's law of scale 1/10 but the presence of gravity showed that the scaling law can be altered to 1/8 due to additional inertial forces via gravity [38]. In our case, the drop is under a forced vibration which is in other words, can be treated as an effective time-varying gravity. Hence, it is important to understand whether such time varying gravity alter the universal scaling law. Having said that, it is evident that the three forces along with the inertial forces via oscillation can compete near the pinch-off event and illustrates another open question, i.e., what forces define the regime for the universal pinch-off event under forced vibration.

The constrained instabilities are usually studied by unconstrained counterpart. It limits the utility but can be used as a foundation to understand the physics of the problem. However, neglecting the contact line information and oversimplification of the problem often leads to misinterpretation of the experimental observations. Parametric excitation of a constrained sessile drop forming subharmonic azimuthal waves is one such problem where the dynamic behavior of the drop is oversimplified to planar Faraday waves in the past and the physics behind the drop vibrations is still a baffling problem in drop atomization applications. Next, the pinch-off mechanism in constrained drop vibration is a non-linear effect and is difficult to analyze with linear theories as well as simplified planar Faraday waves. Therefore, in this work, we utilize a Direct Numerical Simulation (DNS) study to investigate two open questions regarding the physics of the drop vibrations, i.e., (a) What causes the azimuthal waves along the contact line of a vibrating sessile drop? and (b) Is the pinch-off mechanism of the vibrating sessile drop universal (i.e., following a universal visco-capillary or inertia-capillary regime) or the mechanism depends on oscillatory effective gravity?

It is for the first time a detailed mechanism of interfacial waves on a vibrating drop and atomization is studied with DNS that proves that the wave on sessile drop is not entirely governed by Faraday waves. Rather, a two-wave dynamical system is induced that leads to Faraday-like subharmonic response near the contact line. Further, we explain that the primary pinch-off in case of low mode excitation is universal and follows visco-capillary regime irrespective of the inertial forces via oscillatory effective gravity and the secondary pinch-off produces satellite drop which do not follow the universal pinch-off scaling law and hence, the inertial forces via effective gravity deviate the pinch-off mechanism.

Having said that, the rest of the paper is organized as follows: The methodology of DNS comprising the problem formulation, governing equations, contact line model, and numerical scheme is discussed in **sec. 2**. Then, we explain the two-wave dynamical system in case of a high mode excited drop to answer the first open problem, i.e., the cause of production of subharmonic azimuthal waves in **sec. 3**. After that, the dynamics of a low mode excited sessile drop is discussed in **sec. 4**, to illustrate the cause of spike formation via crater, the structure of necking and finally the mechanism of primary and secondary pinch-off to answer the second open problem. Finally, the conclusion and the future works on parametric studies on the vibrated sessile drop is presented in **sec. 5**.

## 2. METHODOLOGY

### 2.1. Problem formulation

In this work, a sessile water drop of contact radius $R$ and density $\rho_w$ is surrounded by air of density $\rho_a$ placed on a vibrating solid surface as shown in **Fig. 2(a)**. The vibratory surface is characterized by vertical acceleration amplitude $A$ and frequency $f$. The time-varying acceleration imposed on the drop is $A\sin(\omega t)$, $\omega$ is the angular frequency (equals $2\pi f$). The acceleration due to gravity is in the negative $z$-direction.

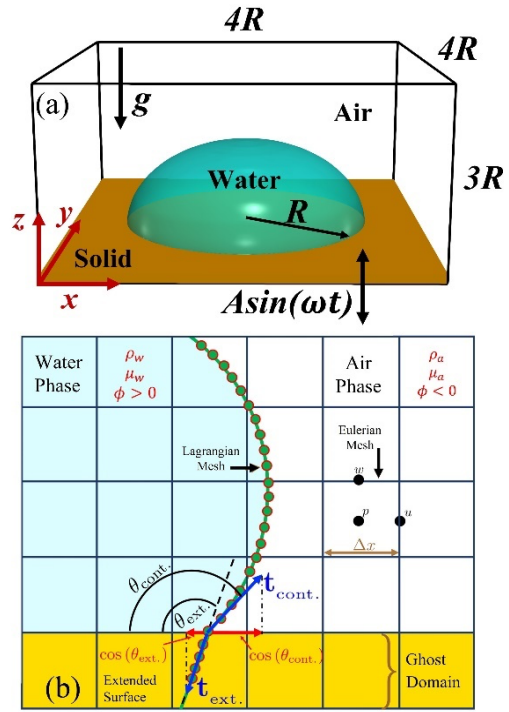

Fig. 2. (a) Problem statement representation: The air, water and solid is visualized by white, cyan and orange colour respectively (b) Two-dimensional illustration of contact-line model in Eulerian-Lagrangian grid with an extended surface

The resultant of the gravity and the vertical vibratory acceleration imposes a time-varying effective gravity in the frame of reference of the sessile drop. The choice of domain size is implemented with the contact radius as the length scale. The $xy$ plane is considered to be $4R \times 4R$, while by assuming the drop ejection should not exceed 2.5R, the height of the domain is considered to be 3R. The boundary conditions are as follow: (a) velocity field: Dirichlet boundary condition at all boundaries except the bottom where no-slip, no-penetration boundary condition is implemented for the solid-fluid interactions, (b) pressure field: Dirichlet boundary condition for all boundaries. The oscillatory time-varying effective gravity provides the body forces which is the source term in the Navier-Stokes equation. Moreover, the dynamics of the contact line is imposed using the Navier-slip condition. In the next sections (i.e., **sec. 2.2, 2.3**), the oscillatory effective gravity and the contact line model for the Navier-slip condition is discussed in detail.

2.2. Governing equations

The governing equation for a single field formulation of incompressible water-air system is expressed by,

$$\nabla \boldsymbol{u} = 0 \quad (2.1(a))$$

$$\rho \left( \frac{\partial \boldsymbol{u}}{\partial t} + \boldsymbol{u} \nabla \boldsymbol{u} \right) = -\nabla P + \nabla \mu (\nabla \boldsymbol{u} + \nabla \boldsymbol{u}^T) + \boldsymbol{G} + \boldsymbol{F} \quad (2.1(b))$$

where, $\boldsymbol{u}$, $P$, $\boldsymbol{F}$, $\boldsymbol{G}$ are the velocity, pressure, local surface tension force at the water-air interface and total homogenous body force respectively. The material properties that distinguish the water-air system are the density $\rho$ and viscosity $\mu$ defined at $\boldsymbol{x} = (x, y, z)$ in the domain by,

$$\rho(\boldsymbol{x}, t) = \rho_a + (\rho_w - \rho_a)\mathcal{H}(\boldsymbol{x}, t) \quad (2.2(a))$$
$$\mu(\boldsymbol{x}, t) = \mu_a + (\mu_w - \mu_a)\mathcal{H}(\boldsymbol{x}, t) \quad (2.2(b))$$

The indicator function $\mathcal{H}(\boldsymbol{x}, t)$ is a numerical Heaviside function which is equal to 0 for the air phase (i.e., subscript $a$) and 1 for the water phase (i.e., subscript $w$) in **Eq. 2.2**. In this work, $\rho_w$, $\rho_a$, $\mu_w$, and $\mu_a$ are equal to 998 kg/m$^3$, 1.205 kg/m$^3$, 1.82×10$^{-5}$ kg/ms, and 10$^{-3}$ kg/ms respectively. The multiphase model utilizes sharp interface system and thus, $\mathcal{H}$ is resolved in a sharp transition of at most 3-4 grid cells across the interface. It is essentially generated by a vector distance function $\phi(\boldsymbol{x})$ computed directly from the tracked interface [39]. Accurate surface tension force calculation plays major role in problems such as atomization where the necking structure for the pinch-off are driven by the dominant surface tension force. In this work, it is implemented by the hybrid formulation following Shin et al. [40] as,

$$\boldsymbol{F} = \sigma K_H \nabla \mathcal{H} \quad (2.3)$$

where, $\sigma$ is the surface tension coefficient (which is constant in this work) (i.e., equal to 0.0714 N/m) and $K_H$ is twice the interface curvature field. It is computed in the Eulerian grid with a time-dependent parameterization of the dynamic interface. The information of the dynamic interface is provided via Lagrangian mesh which is distributed in the Eulerian grid for the computation by following the Peskin's immersed boundary method [41]. For more details regarding the implementation and data transfer between the Eulerian and Lagrangian interface, the readers can refer [39]–[43]. Finally, the term $\boldsymbol{G}$ is the source term which implements the oscillatory acceleration given by effective gravity (i.e., $g - A\sin(\omega t)$) on the drop.

2.3. Contact line model

The solid-air-water is the three-phase contact line which imposes an additional uncompensated Young's stress along with the viscous shear stress on the sessile drop motion. The no-slip boundary condition at the vibrating wall results in an inevitable infinite viscous shear stress near the contact line. The infinite shear stress can be prevented by allowing the contact line to slip along the wall. The contact line slipping is modelled using the Navier slip condition [44] that displaces the interface in proportional to the shear stress exerted near the wall. Yamamoto et al. [45] referred to the modified slipping condition as the Generalized Navier Boundary Condition (GNBC) which tracks accurate contact line motion obtained from molecular theory. GNBC also includes the uncompensated Young's stress which accounts the interaction on a thin layer near the three-phase contact line. Hence, the GNBC is adapted in the current hybrid Front-tracking/Level-set scheme by the Navier-slip condition imposed by,

$$\boldsymbol{U}_{CL} = \lambda \frac{\partial \boldsymbol{u}}{\partial n}\Big|_{wall} \quad (2.4)$$

where, $\boldsymbol{U}_{CL}$ is the contact line velocity, $\partial \boldsymbol{u}/\partial n$ is the shear strain rate and $\lambda$ is the slip length. In theory, the slip length of a contact line is dynamic and hence not constant. However, as explained by Kirkinis and Davis [46], it can be considered to be constant equal to the length of the Eulerian grid ($\Delta x$) in the present study. The implementation details of the three-phase contact line are illustrated using a two-dimensional representation in **Fig. 2(b)**. The solid air and water phase is distinguished as yellow, white and cyan color respectively where the interface is tracked using a Lagrangian mesh (represented in green). The material properties $\rho, \mu$ are computed by the vector distance function $\phi$ such that it is positive for water and negative for air. The dynamic properties $\boldsymbol{u}, p$ are computed in the Eulerian mesh using the numerical procedure described in the next section (i.e., **sec. 2.4**).

An artificial ghost domain for the solid surface is utilized in the Lagrangian-Eulerian mesh to consider the interfacial forces at the three-phase contact line. This ghost domain is the extended surface which is correctly manipulated to account the Navier slip condition (**Eq. 2.4**) as shown in **Fig. 2(b)**. Using the interface normal to the element contact the solid wall ($t_{cont.}$) and a vertical normal to the solid wall from $t_{cont}$ (see the dotted line from $t_{cont}$), the contact angle $\theta_{cont}$ is evaluated. Similarly, $\theta_{ext.}$ is evaluated in such a way that it accurately tracks the interfacial forces accounting the viscous as well as Young's stresses near the wall. The choice of the $\theta_{ext.}$ depends on the contact angle hysteresis using the hysteresis model given by,

$$\begin{cases} \theta_{ext.} = \theta_{adv} \text{ ; if } \theta_{cont.} > \theta_{adv} \\ \theta_{ext.} = \theta_{rec} \text{ ; if } \theta_{cont.} < \theta_{rec} \\ \theta_{adv} < \theta_{ext.} < \theta_{rec} \text{ ; otherwise} \end{cases} \quad (2.5)$$

where, the $\theta_{adv}$ and $\theta_{rec}$ are the advancing and receding contact angle respectively.

## 2.4. Numerical scheme

A conventional projection method on a staggered mesh is implemented where the **Eq. 2.1** is discretized as,

$$\frac{\boldsymbol{u}^{n+1} - \boldsymbol{u}^n}{\Delta t} = \frac{1}{\rho^n}(-\boldsymbol{u}.\boldsymbol{\nabla}\boldsymbol{u})^n + \boldsymbol{G}^n + \boldsymbol{F}^n) - \frac{1}{\rho^n}\nabla_h P \quad (2.6)$$

where, $h$ is the spatially discrete operator. Second-order essentially non-oscillatory (ENO) method is utilized for the convection term and central difference method is utilized for the diffusion term. Two sub-steps are implemented for the time integration procedure wherein semi-implicit calculation of an intermediate unprojected velocity ($\boldsymbol{u}'$) is first performed by ignoring pressure and involving only the velocities and their gradients. Then, by enforcing the divergence-free property of the velocity, the pressure term is calculated. Finally, the updated velocity $\boldsymbol{u}^{n+1}$ is obtained by,

$$\boldsymbol{u}^{n+1} = \boldsymbol{u}' - \frac{\Delta t}{\rho^n}\nabla_h P \quad (2.7)$$

The temporal iteration time step $\Delta t$ is chosen in such a way that it satisfies the minimization criterion based on,

$$\Delta t = \min(\Delta t_{CFL}, \Delta t_{int}, \Delta t_{vis}, \Delta t_{cap}, \Delta t_\omega) \quad (2.8)$$

The $\Delta t_{CFL}, \Delta t_{int}, \Delta t_{vis}, \Delta t_{cap}$, are the Courant–Friedrichs–Lewy (CFL), inertial, viscous, and capillary time steps respectively. The definitions of these time steps are followed from [47] and the $\Delta t_\omega$ is the oscillatory time step which supplements an upper bound equal to $2\pi/50\omega$. The calculations of the given problem statement is performed using FORTRAN 2003. The code is highly parallelized based on domain decomposition procedures by a parallel GMRES method to calculate implicit viscous terms for the velocity field and by a parallel multigrid algorithm for the pressure calculations. Communication across processing slaves is handled by MPI procedures. Finally, the code contains a module for the immersed solid objects and their interactions with the water-air system. The Navier-slip dynamic contact line model is implemented within the module for the problem as prescribed by Shin and Juric [43].

We utilized two domains for high mode and low mode excitation in **sec. 3** and **sec. 4** respectively. In both the cases, we utilize a uniform mesh where each Eulerian grid is cubic and each parallel core in the code handles $32^3$ mesh grids. The code is utilized in previous works with oscillatory body forces where it concluded that for a mesh independent calculation, the wavelength of the interface must be resolved accurately. Henceforth, we utilize finer resolution in case of high mode excitation (where the wavelength is smaller than the low mode excitation). In case of low mode excitation, we implemented 8×8×6 parallel cores (i.e., $1.25 \times 10^7$ mesh grids) while in the case of high mode excitation, 12×12×6 parallel cores (i.e., $2.83 \times 10^7$ mesh grids). As a final note, it is informed that the effect of air on the vibrating drop is negligible, and the remainder of the paper considers water as a fluid (or aliased as liquid) to prevent confusion for the reader.

## 3. HIGH MODE EXCITATION

In this section, we present the explanation of the first open problem, i.e., the mechanism of the subharmonic response in vibration of sessile drops. Vukasinovic et al. [24] are one of the first researchers who demonstrated the subharmonic response on a sessile drop under high mode of excitation. Inspired from the experimental set-up, we demonstrate our simulation in the similar procedure. First, a sessile drop of 100 $\mu L$ is placed on a vibrating diaphragm at 1040 kHz. The contact radius of the drop is 4.2 mm and assuming that the drop is flattened due to gravity, the vertical radius obtained is 2.7 mm. The static contact angle is equal to 84° and the advancing and receding angles are 87° and 81° respectively. Such a high frequency inevitably showcases a high mode of excitation. It is interesting to note that the simulation of such sessile drop vibration is not studied in any of the previous works, and it is for the first time we are demonstrating it in the fluid dynamics community. First, the acceleration amplitude is set to a lower value (200 m/s$^2$ in our case) to initiate the harmonics on the drop, i.e., axisymmetric oscillation. The characteristic wavelength of such a vibration of the solid is 26 $\mu m$ and thus, initiates small ripples along the contact line as soon as vibration begins. This observation thus resembles a damped spring mass system of single degree of freedom. After that, the acceleration is slowly ramped to 1000 m/s$^2$ where the superposition of the axisymmetric and azimuthal waves is observed. The characteristic wavelength of such system is 24.3 $\mu m$. As the characteristic wavelength of the vibration in both cases is negligible in comparison to the sessile drop's radius, the rapid deformation of the interface by virtue of drop's inertia is not observed in the case of high mode excited sessile drop vibration. On the other hand, in **sec. 4**, a characteristic wavelength of similar order initiates drop's ejection and rich dynamics of pinch-off (which is discussed in detail). Having said that, the two wave modes via axisymmetric and azimuthal response is discussed in detail in the next two subsection (i.e., **sec. 3.1, 3.2**) and their qualitative validation is provided in **Appendix. A.2**.

### 3.1. Axisymmetric standing waves

Driving the vibration at 200 m/s$^2$ excites the sessile drop in the high mode regime. **Fig. 3(a)** represents the interfacial position from t = 44T (where T refers to time period in the remainder of the paper) to 44T + 7T/4 (at an interval of T/4) to describe the dynamics of the wave with time. It is observed that the position of the interface is equivalent after each T. For example, the interfacial position at t = 44T and t + T are strikingly coinciding and thus validates that the waves are harmonic in nature. Moreover, as the amplitude of the interfacial displacement remain constant with time (i.e., the maximum deviation at t + T/4 and t + 5T/4 are equal), the resultant waves are proved to be standing waves. It is interesting to observe that the maximum deviation at the top of the sessile drop and minimum near the contact line. The heterogeneity of the interfacial wave amplitude across the interface is due to the poloidal dependence of the fluctuation to the forced vibrating acceleration as discussed in **sec. 1**. These poloidal dependences of fluctuation have two extremities, i.e., (i) the tip of the sessile drop (topmost region) where the vertical vibration is parallel to the normal of the interface and hence, resembles Faraday waves, and (ii) the contact line of the sessile drop (lowermost region) where the vertical vibration is almost perpendicular to the normal of the interface showcasing oscillatory Kelvin-Helmholtz like waves.

Not only the interfacial fluctuations, but also the fluid flow regimes are observed to follow similar axisymmetric standing harmonic waves as shown in **Fig. 3(b, c)**. The interfacial pressure fluctuations are plotted in **Fig. 3(b)** from t = 44T to t + 7T/4 and the pressure contours at t = 44T are shown in **Fig. 3(c)**. A qualitative argument can be drawn from the pressure contours where axisymmetric concentric waves are observed whereas the pressure fluctuation plot proves that the interfacial pressure is also axisymmetric, harmonic and standing wave.

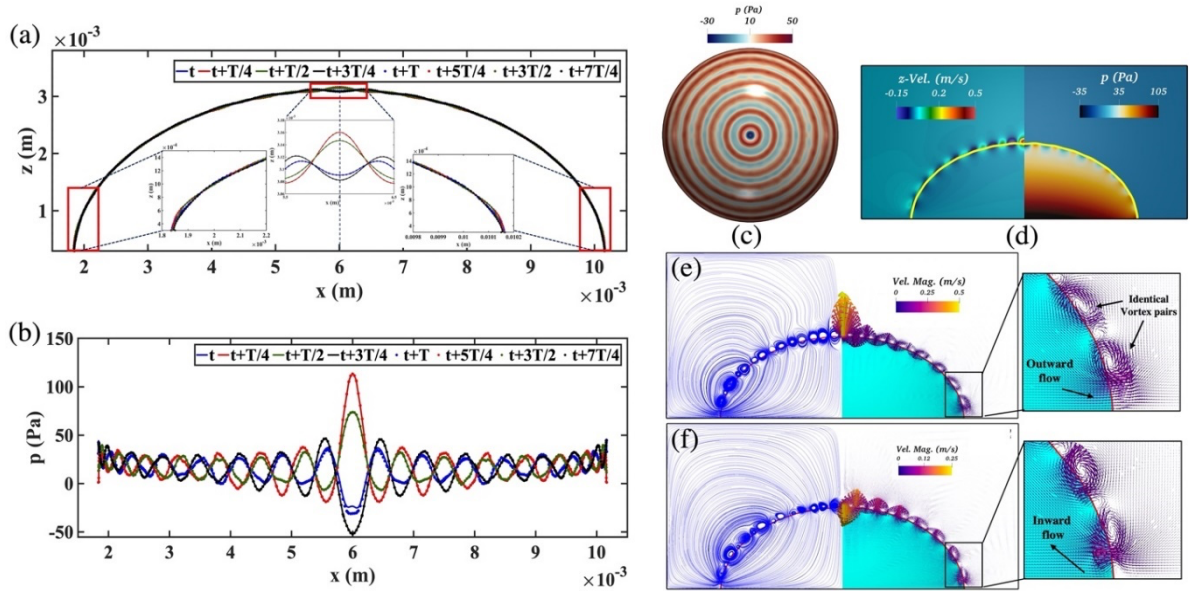

Fig. 3. (a) Interfacial fluctuation and (b) pressure fields from t=44T to t+7T/4 at an interval of T/4 (c) pressure contours on the midsection at t = 44T ($xy$ plane) (d) $z$-velocity and pressure contours at mid-section across the sessile drop ($xz$ plane) (e-f) velocity streamlines and glyphs at t = 44T and t + T/2 respectively

Similar analysis for velocity field is shown in **Appendix. A.1**. From **Fig. 3(b)**, the number of waves (in one half) formed are 8 which correspond to a mode-8 harmonic standing wave of a mean wavelength equal to 1.57 mm (2.5x lower than the sessile drop radius). Additionally, as the waves are harmonic, the frequency of the concentric waves are equal to 1040 Hz and the wave speed of such waves is equal to 2.66 m/s. Therefore, it is evident to observe 8 concentric crests as well as troughs on the interface as shown in the pressure contours in **Fig. 3(c)**. Moreover, the $z$-velocity and pressure fields at the longitudinal midsection of the sessile drop ($xz$ plane) signifies that the interfacial dynamics are dominant over the bulk fluid dynamics as the characteristic wavelength of the vibrating solid is 100 times smaller than the sessile drop radius. Hence, the interfacial dynamics dominate the waves in this regime as shown in **Fig. 3(d)**. The velocity streamlines and glyphs showcase the vortex interaction across the interface in **Fig. 3(e, f)** at t = 44T and t + T/2 respectively. It is interesting to observe that at t = 44T, the identical vortices induce the flow outward (see the inset of **Fig. 3(e)** near the contact line) which consequently oppose the maximum velocity attained at the tip of the sessile drop. Next, at half cycle (i.e., t + T/2), the velocity at the tip changes its direction to negative $z$-direction. At this interval, the identical vortices change its direction in such a way that the flow is induced inward (see the inset of the glyphs in **Fig. 3(f)** near the contact line) to oppose the maximum velocity attained at the tip. These fluxes (either inward or outward) oppose and affect the dynamics at the tip of the sessile drop. Moreover, the maximum of these fluxes occurs near the contact line due to the lowest poloidal angle dependence as shown qualitatively in the streamlines and glyphs of **Fig. 3(e, f)**. Therefore, a two-wave dynamical system can be determined between the two extremities, i.e, the tip and the contact line of the sessile drop. In other words, the flow induced by the waves along the contact line opposes the dynamics at the tip of the sessile drop. It proves that rather than the dependence of the forced vibration of the solid, the varied dynamics near the contact line and the tip of the sessile are interrelated and forms a *self-induced* dynamical system. The relationship between the waves generated along the contact line and the tip of the sessile drop becomes prominent as the acceleration increases and hence, we observe the azimuthal waves. Further discussion on the formation of azimuthal waves and its deliberate analysis is presented in **sec. 3.2**. It is to be noted to the reader that the dynamics *along* the contact line doesn't alias with the three-phase line rather it refers to the region *near* the contact line which deforms under vibration within the limits of advancing and receding angle.

### 3.2. Azimuthal wave

The process of generating the azimuthal waves on the sessile drop is not as straightforward as producing the axisymmetric waves. The axisymmetric waves are found to be harmonic in nature, meaning the interfacial waves are produced at the same frequency, i.e, 1040 Hz as of the vibrating solid. The mode-8 standing vibration can be obtained at the onset of the instability following the same phenomenon as a forced pendulum at the given frequency or a linear damped spring-mass system forced under constant frequency. The case of subharmonic response is a classical counter-intuitive case most commonly obtained via parametric oscillation of an inverted pendulum, famously called the Kapitsa pendulum. In such cases, the pendulum is under a parametric frequency near to its resonance regime, but the frequency observed by the pendulum is *half* of the parametric frequency (or twice the tome period T). It is easier to obtain such subharmonic regime on flat fluid interface under vertical vibration with no vital role of the solid-fluid interactions. However, in the present case, the contact line is also under oscillatory vibration. In order to obtain the subharmonic response, the sessile drop is needed to go through a series of increasing acceleration amplitude to sweep through maximum mode of vibration possible for the sessile drop.

In our case, the sessile drop went through at least 25 time periods (T) for each acceleration amplitude at an interval of 100 m/s$^2$ from 200 m/s$^2$ to 1000 m/s$^2$. In this process, the amplitude of drop vibration also increases and thus, affects the self-induced dynamical system between the tip and the contact line of the sessile drop. **Fig. 4(a, b)** showcases that at t = 512T + T/2, dimple shaped high pressure zones creating new crests along the contact line of the sessile drop. These dimple shaped zones prove the presence of azimuthal waves along the contact line of the drop. It is observed that there are 21 dimples (crests) along the circumference of the contact line. Hence, the azimuthal wavelength is found to be 1.25 mm which is comparable but lesser than the axisymmetric wavelength of 1.57 mm.

In order to prove that the azimuthal waves are subharmonic response, the pressure contours, radial velocity streamlines, glyphs and contours for the $xy$ plane at the depth of the contact line are shown in quadrant I, II, III, and IV at t = 512T + T/2, t + T, and t + 2T are shown in **Fig. 4(c-e)** respectively. A diagonal line is drawn in the quadrant I and IV for the pressure contours as well as the radial velocity glyphs in the inset to characterize the wave dynamics. At t = 512T + T/2, the line intersects with a dimple shaped crest on the interface which is at higher pressure. At one full cycle (i.e., t + T), it is observed that the line passes through two dimple shaped high pressure zones featuring a trough at the interface. Finally, in the second cycle (i.e., t + 2T), the line passes through the high pressure dimpled shaped zone again proving that the azimuthal waves complete one full cycle (i.e., from one crest to another) in 2T time period of the force oscillation. In other words, the azimuthal waves are under an oscillation equal to half of the driving frequency and therefore, it proves that the high-mode vibration of the sessile drops produce azimuthal waves with subharmonic response. Similar argument can also be drawn from the velocity glyphs as shown in the inset of **Fig. 4(c-e)**, where the high radial velocity is attained at the troughs.

It is interesting to observe that the fluid flow is in radial direction as the velocity glyphs are radially inward at every full cycle T. Thus, the radial flow suggests that the interfacial waves are governed by Faraday-like instabilities where the oscillating fluid flow should be in the radial direction as discussed in **sec. 1**. However, these Faraday-like instabilities are not caused by the harmonics provided via vertical vibration of the solid surface as discussed in **sec. 1**. Henceforth, there must be another source of vibration by which such radially directed Faraday-like deformations are formed.

The other source of vibration than the vertical solid vibration is the interfacial waves itself where the axisymmetric standing waves are observed to be coupled with the azimuthal waves. These interfacial waves are harmonic, i.e., waves generated are on the same frequency of the vibrating solid. Unlike the axisymmetric waves case where the dynamics of the tip and the contact line are in out-of-phase, i.e., the dynamics opposes the effect of other (in simpler words, when the flow at the tip is inward, the flow near the contact line is also inward to oppose the effect at the tip), the azimuthal case shows in-phase dynamics where the tip and the contact line fluid flow do not oppose each other due to high mode of acceleration. To analyze such self-induced system between the tip and the contact line, the velocity streamlines and glyphs at a half cycle apart (i.e., t = 512T + T/2 and t + T/2) are shown in **Fig. 4(f, g)**. At t = 512T + T/2, the fluid influx across the contact line aids the velocity at the tip of the sessile drop as shown in **Fig. 4(f)** and at t + T/2, the tip velocity aids the fluid outflux across the contact line (see **Fig. 4(g)**). Thus, the in-phase self-induced system shows that the trough in the azimuthal waves provide the necessary inward flow for the maximum upward velocity to the tip of the sessile drop and the downward velocity at the tip at the next half cycle aids the deformation of the trough to become a crest via outward flow near the contact line.

The mechanism of the subharmonic response is as follow: Let say there is a trough in the azimuthal direction near the contact line providing sufficient inward flux at t = 512T + T/2 as shown in **Fig. 4(c, f)**. At this point, the standing harmonic waves at the tip starts to decrease in magnitude and as it is in-phase with contact line, the inward velocity of the trough near the contact line also decreases which consequently move the trough to its mean position. At t + T/2, the velocity at the tip changes its direction towards negative *z*-direction, which aids an outward flow at the contact line- further pushes the trough to become crest. The trough position thus becomes a crest at t + T interval (i.e., inward velocity as shown in **Fig. 4(d)**). At this point, the adjacent sides of the formed crest become new troughs (refer **Fig. 4(d)**), and the same phenomenon occurs for the next one cycle to come back to its trough position as shown in **Fig. 4(e)**. Hence, the harmonic oscillations of the vibrating fluid interface itself generates the subharmonic azimuthal waves near the contact line. **Fig. 4(h)** summarizes the dynamical system where the dynamics near the tip affects the dynamics near the contact line of the sessile drop and vice versa. In case of axisymmetric waves, the flow at both extremities opposes each other, whereas in case of azimuthal, the flows aid the dynamics at the extremities. At t, one trough near the contact line aids the dynamical system and at t + T, the adjacent newly formed troughs again aid the flow at the tip. Such a dynamical system is analogous to a parametric driven double pendulum system where the harmonic modes of larger mass constitute the subharmonic modes in the smaller mass. Sarkar et al. [48] showcased that such dynamical system shows rich dynamics and there are several possible configurations at which the double pendulum shows overlapped harmonic and subharmonic response similar to the present observation on a vibrated sessile drop. Hence, it is concluded that even though the waves are Faraday-like instabilities, it is not produced directly via the vertically vibrating solid surface harmonics. Rather, it is the interfacial harmonic waves itself that induce the radially directed parametric oscillations for the contact line in the azimuthal direction.

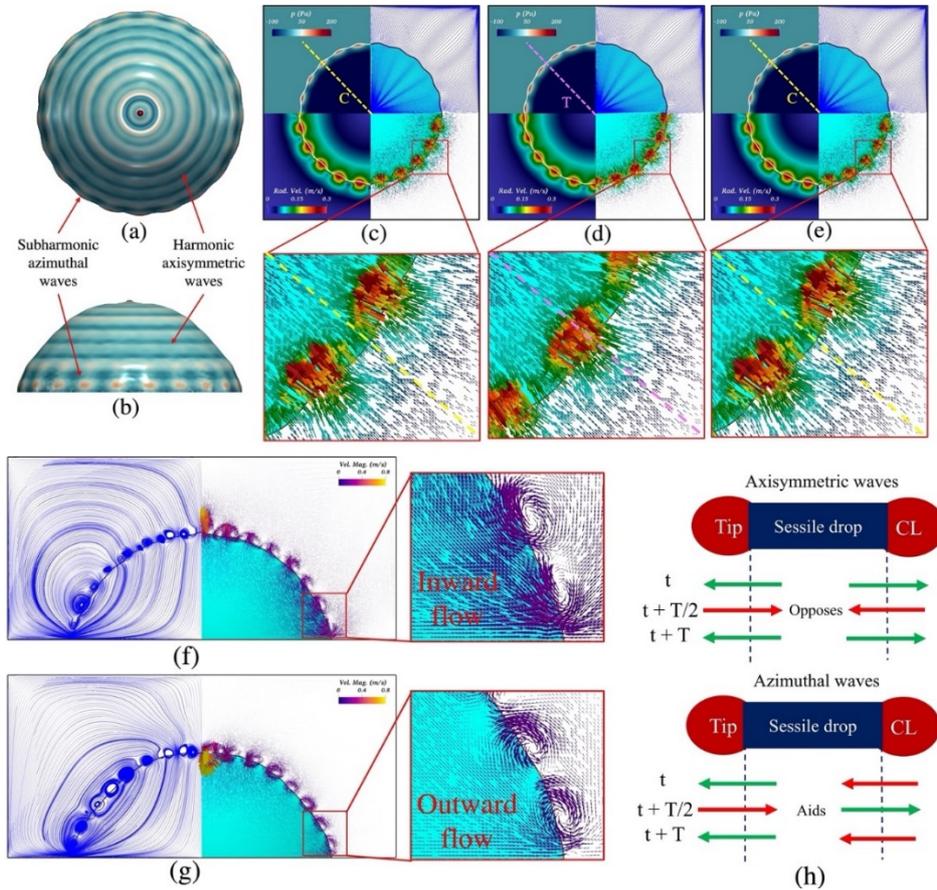

Fig. 4. Pressure contours visualization at t = 512T+T/2 from (a) $xy$ and (b) $yz$ plane for azimuthal waves along the contact line of the sessile drop. Slice across the contact line representing pressure contours, radial velocity streamlines, glyphs and contours in quadrant I, II, III and IV at (c) t = 512T+T/2, (d) t + T, and (e) t + 2T. Velocity streamlines and glyphs in $yz$ plane at (f) t = 512T+T/2 and (g) t + T/2   (Note: the C refers to "Crest" and the T refers to "Trough" in (c-e) while it refers to time period in the remainder of the paper) (h) Tip-contact line (CL) dynamical system

The analysis at high mode excitation showed an interesting observation in both axisymmetric and the azimuthal waves, i.e., the maximum velocity is always attained at the tip of the sessile drop and hence the maximum deviations are also observed at the tip. Therefore, at a threshold acceleration, pinch-off and further bursting dynamics from the surface is inevitable. Upon increasing the acceleration, the azimuthal waves begin to couple with the axisymmetric waves and grows towards the tip of the sessile drop and break the symmetry in the drop. However, forcing such a high acceleration means sweeping higher modes of oscillation and hence smaller wavelength. As the wavelength decreases, the resolution of the mesh gets finer and the computational cost even for one T increases in folds. Therefore, we choose a low mode excitation where the characteristic wavelength of the parametric vibration is in the order of the sessile drop's radius and thus, the atomization is attained in reasonable computational cost.

## 4. LOW MODE EXCITATION

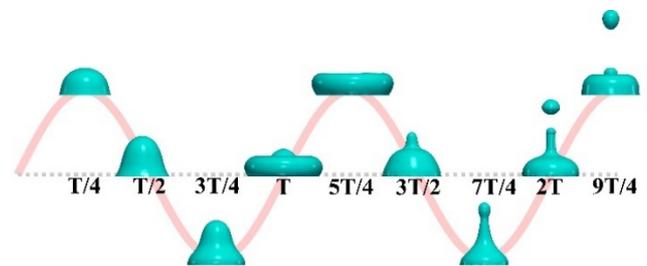

Fig. 5. Sessile drop vibration and atomization for a 2T time- prominent features include (i) elongation of a drop from t=0 to 3T/4, (ii) crater formation from t=3T/4 to 5T/4, (iii) conical spike formation from t=5T/4 to 7T/4, (iv) primary capillary pinch-off, ligament retraction and secondary capillary pinch-off from t=7T/4 to 9T/4 (rich dynamics in short interval: the analysis is discussed aliased to seconds)

In this section, the focus of the work is switched to understanding the dynamics of the atomization of drops under vertical excitation. For the purpose of analysis, a smaller drop excited at relatively lower frequency and amplitude relative to the previous section. The choice of lower frequency enables a longer wavelength that reduces the computational cost as coarser resolution can be utilized as discussed in **sec. 2.4**. Additionally, a gentle amplitude of acceleration comparable to the low frequency of vibration provides enough agitation and negative interfacial curvature for the atomization process. Therefore, inspired by the experimental set-up of James et al. [49], we choose a drop of 30 $\mu L$ under a periodic vertical vibration of amplitude $A$ equals 66 m/s$^2$ and frequency $f$ equals 61 Hz. Assuming the gravity effects do not flatten the interface initially as the drop is small, the initial shape of the drop is considered to be a hemisphere. The contact angle is therefore, 90° and the advancing and receding angle are neglected as the effects of the tip and contact line is overcome by the drop's bulk inertia under low mode excitation. The contact radius is set to 2.5 mm. Furthermore, the characteristic wavelength is obtained to be 2.8 mm. As the characteristic wavelength of the wave generated due to the vertical vibration is greater than the radius of the drop, mode-1 excitation is obtained under such parameters. It implies that single mode wave characteristics are observed. Further, the derived wave characteristics of the forced vibrations are the amplitude length and amplitude velocity equal to 0.44 mm and 17.2 mm/s respectively. It implies that the action (amplitude velocity) and the vertical displacement (amplitude length) of the forced vibration are in the order of 10 and 10$^{-2}$ with respect to the drop's radius respectively. The wide range of reaction and displacement of the wave parameters on the drop leads to distinctive lag behavior of the sessile drop due to inertial effects. It is thus beneficial to utilize these parameters to understand the dynamics of drop atomization under sessile drop excitation as single mode crest and troughs provide the necessary condition for the crater formation which plays a major role in the atomization process.

The atomization process is described in brevity as follow: First, the vibrating surface moves up from t = 0 to t = T/4 and the sessile drop undergoes flattening of the spherical cap due to drop's inertia at the tip of the drop (see **Fig. 5** at t = T/4). In the next half cycle (i.e., t = T/4 to 3T/4), the vibrating surface moves down with an increasing acceleration. The acceleration with a higher amplitude velocity forms an inertial zone for the drop where the bottom surface moves under the reaction of the vibrating surface while the top half remains intact due to lag behavior of drop's inertia. It leads to the formation of an elongated drop as shown in **Fig. 5** at t = 3T/4. Further half cycle (i.e., t = 3T/4 to 5T/4) can also be elucidated using the similar argument but in opposite direction. Here, the peripheral zone near to the contact line uplifts with a decreasing acceleration upon which it moves up while the central zone possess the inertial lag behavior leading to the formation of a crater (see **Fig. 5** at t = 5T/4). A crater is a depression providing necessary negative interfacial curvature to emanate local high pressure zones leading to a sudden conical spike formation. The conical spike formed due to the local pressure zones moves upwards with a high velocity. The inertial body forces become negligible as soon as it reaches the trough position. In this cycle, the surface tension forces become dominant and cause the bulbous structure at the tip of the spike as shown in **Fig. 5** at t ≈ 7T/4. Such a bulbous structure forms the necking event in the spike which further detaches and hence, the occurrence of 'primary' pinch-off mechanism is observed. In the last half cycle, as soon as the pinch-off occurs, a retracting ligament is formed (see **Fig. 5** at t ≈ 2T). The lag behavior of the drop's inertia and the oscillating body forces provide enough dominance of the surface tension leading to the formation of another bulbous structure near to the central zone of the drop. This leads to a 'secondary' pinch-off mechanism and the detached drop then falls back and coalesce in the vibrating drop. The qualitative validation of the simulation provided in **Appendix. A.2**.

The above explanation of the dynamics of the drop vibration under low frequency excitation provide 5 prime events, i.e., crater and spike formation, primary pinch-off, ligament retraction, and secondary pinch-off. In the following sections, the quantitative and qualitative investigation of each event is discussed in detail.

### 4.1. Crater and spike formation

As discussed in the previous section, the drop's inertia lag while an upward motion of the contact line leads to the formation of a negative curvature interface at the central region commonly called the crater formation. In this section, we discuss the dynamics of the crater formation and its consequence, i.e., the formation of a conical spike.

**Fig. 6** represents two snapshots for velocity magnitude, pressure, velocity streamlines and glyphs at t = 1.18T (0.0195s) and t = 1.34T (0.0220 s) respectively to elucidate the crater structure as well as the dynamics of the drop's interface. In **Fig. 6(a)**, it is observed that at 0.0195s, a negative pressure of -130 Pa is induced at the crater structure due to the negative curvature of the interface. Further, the crater flattens at the central region and creates nearly a zero-pressure zone at the center. On the other hand, the peripheral region induces a high pressure zone (i.e., 150 Pa) with positive curvature. Due to the heterogeneity of the structure and dynamics of the curvature at the center and the periphery, a stationary zone is expected in the bulk where the bulk liquid flows from the center and from the periphery in two different directions. Hence, two vortices of opposite signs, (i.e., V1 and V2) are observed as shown in the streamlines and glyphs of **Fig. 6(a)**. V1 causes the radially outward flow while the V2 causes the axially upward flow at this time interval. At t = 0.0220 s, the peripheral region reaches its maximum curvature as well as the crater reaches its maximum depth, and the central region is no longer flat. Henceforth, the zero-pressure zone at the center collapses and the negative curvature (almost equivalent to an inverted conical section) produces a high negative pressure zone (i.e., -240 Pa) (refer **Fig. 6(b)**). On the other hand, the rate of curvature decreases at the periphery leading to a pressure zone of 120 Pa (lesser than 150 Pa at t = 0.0195 s). Such a large pressure gradient from the periphery to the central region induces a momentum flux towards the center. At this time interval, the vortex V1 also collapses due to no flat interface at the center and only V2 dominates the radial flow. The V2 changes its orientation due to change in the curvature of the periphery and now causes a radially inward flow from the periphery to the center as well as an uplifting flow at the center of the crater (see the glyphs in **Fig. 6(b)**). The V2 thus causes the formation of the conical spike from the bottom of the crater.

**Fig. 7(a)** shows the velocity magnitude and pressure contours, the velocity streamlines and glyphs of a snapshot at t = 0.0266 s. It is interesting to note that the velocity magnitude is maximum at the tip of the spike and drops spatially to nearly to zero below a depth of 2.2 mm. The velocity drop is in good agreement with unconstrained vertical accelerated instabilities of Faraday waves by Umemera et al. [50]. They concluded that the effect of the interfacial deformation is significant upto a depth of 1 wavelength order from the interface (in this case, it is 2.8 mm). The pressure contours signify that the maximum pressure is at the tip of the spike due to the initial radial bulk momentum towards the center of the crater. Moreover, due to the negative curvature of the conical spike, a minimal negative pressure zone is observed. The negative pressure zone creates local high-pressure points (see P1 and P2) which differentiates the momentum flux in the bulk and the conical spike. Additionally, the negative curvature dictates the V2 which was earlier dictated by a positive curvature. Henceforth, the strength of V2 decreases and thus, the velocity magnitude decreases with time.

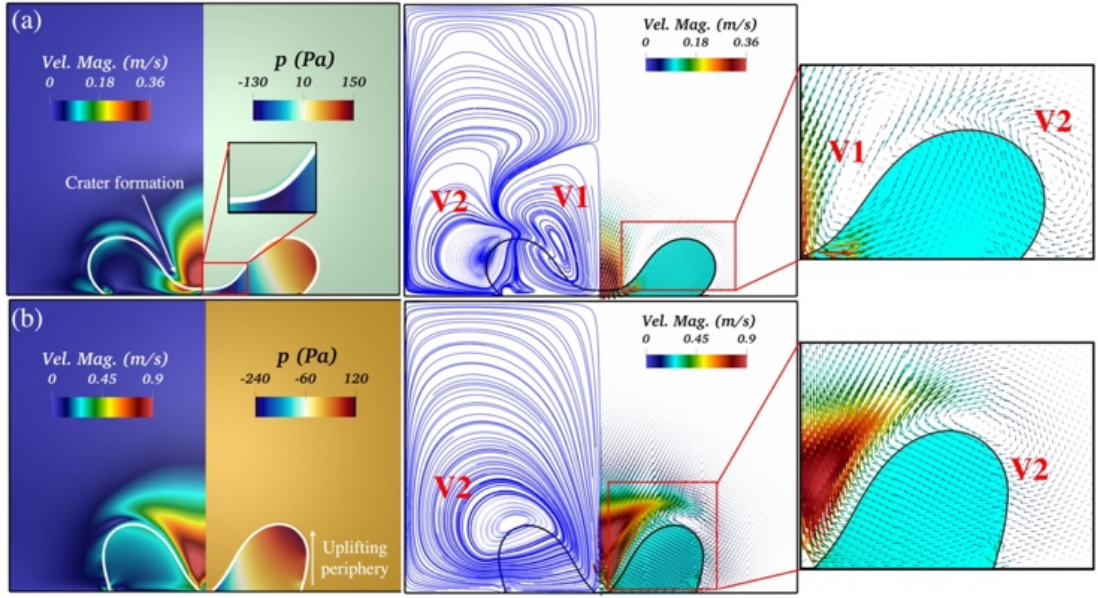

Fig. 6. Crater structure analysis: (left) Velocity magnitude (Vel. Mag.) and pressure contours (p) (right) velocity streamlines and glyphs at (a) 1.18T (0.0195 s) and (b) 1.34T (0.0220 s)

**Fig. 7(b)** represents the decrement of the velocity magnitude with time due to the weakening of the vortex V2 as well as an opposite direction of the pressure gradient (due to the maximum pressure zone at the tip). The initial upward velocity magnitude at t = 0.0225 s is 4.04 m/s and the final tip velocity at the onset of primary pinch-off is almost 0, i.e., 0.16 m/s. It is beneficial to understand the dynamics of the conical spike in terms of three competing forces, i.e., inertial forces mainly via oscillatory effective gravity forces, damping viscous forces and the surface tension forces. One of the ways to delineate the dynamics is via linear stability analysis. For example, the capillary-gravity waves and their dispersion relation provide necessary condition for the competing surface tension and inertial forces via gravity. In case of forced vibration, **Eq. 4.1** is modified by the natural frequency of the fluid obtained via capillary-gravity dispersion relations to obtain the Mathieu equation. However, in the present case, such linear stability analysis cannot be utilized as the length scale of deformation is comparable to the characteristic wavelength. Therefore, a systematic separable strategy is implemented where an ideal dynamical system first can be written as,

$$\frac{d^2L}{dt^2} + (g - A\sin(\omega t)) = 0 \quad (4.1)$$

$L$ is the length of the conical spike and the viscous and surface tension forces are neglected. Moreover, it is assumed that the dynamics of the spike is 1D in the upward direction (z-axis) and no other inertial force e.g., via pressure gradient is there in the system. The initial condition for the dynamical system is given by the height and velocity magnitude of the spike at t = 0.0225 s, i.e., 0.38 mm and 4.04 m/s respectively. The initial value problem is then solved using ODE45 in MATLAB. It is observed that the ideal system showcases an insignificant drop in the velocity magnitude and then begins to increase with time. It infers that the inertial body forces play a role (but not significant) in the initial decrement of the velocity magnitude. Next, the viscous forces are included to the 1D dynamical system by introducing a damping factor $\gamma$ given by,

$$\frac{d^2L}{dt^2} + 2\gamma\frac{dL}{dt} + (g - A\sin(\omega t)) = 0 \quad (4.2)$$

The damping factor is obtained by following the argument by Landau and Lifshitz [51] and Kumar et al. [52] as,

$$\gamma = 2k^2 \frac{\mu_w + \mu_a}{\rho_w + \rho_a} \qquad (4.3)$$

Where, $k$ is the wavenumber defined by the reciprocal of the characteristic wavelength of 2.8 mm. The viscous damped case improved the model as unlike the ideal case, the velocity magnitude is observed to decrease at all time intervals. However, rate of decrement is still not in agreement with the numerical simulation of the present work. Further, three cases are considered where the fluid is highly viscous (i.e., 10x, 11x, 12x compared to water) in such a way that the surface tension forces are negligible. In such case, the initial decrement of the velocity magnitude is found to be in good agreement with the numerical simulation. The 10x viscous fluid fits the final time interval (the necking stage) of the tip velocity while the 12x viscous fluid fits the initial time interval (spike stage) of the tip velocity. The 11x viscous fluid is unbiased to the fitting and can be considered as the best analogous viscous fluid model to the present work.

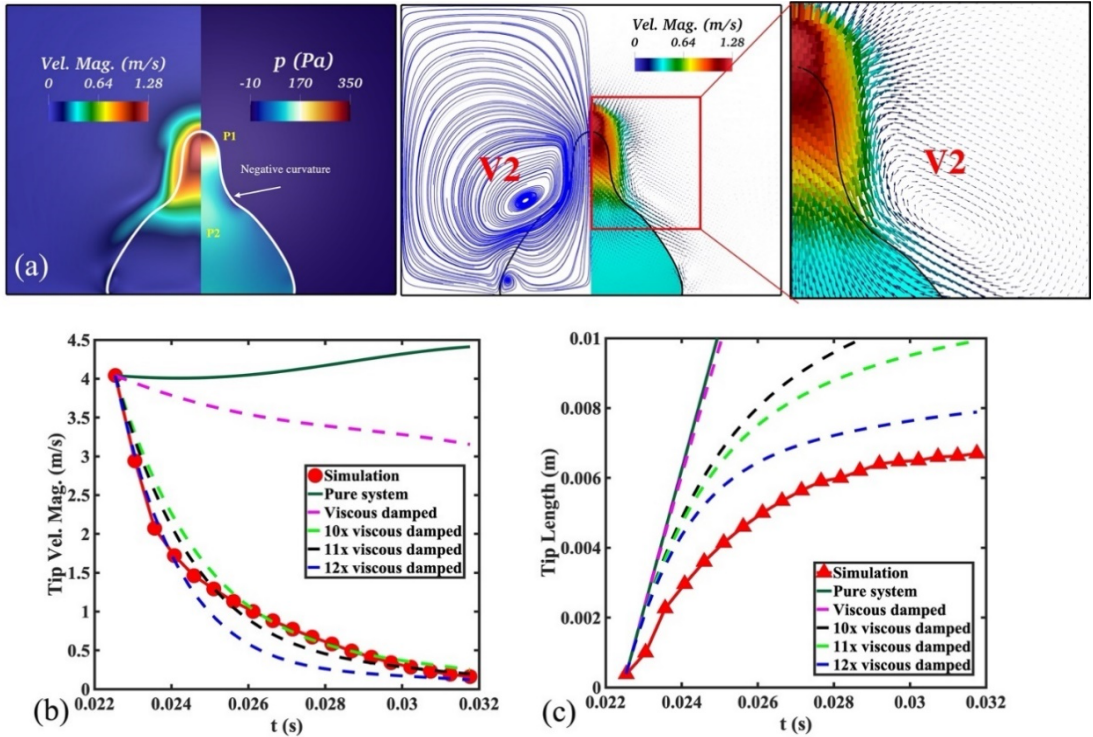

Fig. 7. Conical spike analysis: (left) Velocity magnitude (Vel. Mag.) and pressure contours (p) (right) velocity streamlines and glyphs at 1.625T (0.0266 s) (b) Tip velocity magnitude analysis over time- Prominent events: (i) conical spike tip velocity significantly drops (higher than the ideal and damped dynamical system) (ii) tip velocity drops linearly once at the onset of necking phenomenon (c) Tip length analysis over time- Prominent events: (i) conical spike tip length shoots initially but lower slope than the ideal and damped dynamical system (ii) rate of tip length increment decreases significantly compared to ideal and damped systems

**Fig. 7(c)** represents the tip length dynamics with respect to time interval. As a response to the initial high velocity, the tip length is first increasing rapidly (in the spike stage) and then flattens out with the progress of time (in the necking stage). The ideal and viscous dynamical system are compared with the numerical simulation as shown in **Fig. 7(c)**. The ideal and the viscous damped model are not in good agreement with the dynamics of the tip length in the present study because of significant differences in the velocity magnitude with the numerical simulation. Additionally, it is found that the three highly viscous fluid qualitatively track the dynamics of the dynamics of tip length. The 12x viscous damped fluid showcases the best fit as the initial slope of the velocity magnitude is in good fit with the numerical simulation. Hence, the initial response of the tip length is in order of the 12x viscous fluid.

The above analyses conclude that the oscillatory inertial and viscous forces do not dominate the dynamics of the conical spike. Hence, the surface tension forces are the dominant force which dictate the rapid fall-off of the velocity magnitude and the formation of necking region at t = 0.0272 s. On the contrary, in cases of highly viscous fluids such as the 10x, 11x, and 12x viscous fluids, even though the tip tracks the qualitative and quantitative dynamics of the conical spike, it can not form the necking region and thus, the bulbous structure leading to the pinch-off mechanism for atomization is not possible. Such fluids shoot to similar height and beyond and then drops with respect to the forced vibration. These observations are in agreement with Vukasinovic et al. [53], where they considered two fluids, i.e., water and glycerin-water solution (18x viscous than water) and found that the surface tension delays as the viscous forces prevent the early necking. By concluding that the surface tension forces cause the necking phenomenon, irrespective of the oscillatory behavior. Having said that, the pinch-off mechanism and its dynamics are further elucidated in the next section (i.e., **sec. 4.2**).

### 4.2. Primary pinch-off

The dynamics of the pinch-off event is elucidated by 4 snapshots at t = 0.0271, 0.0297, 0.0312, and 0.0318 s respectively as shown in **Fig. 8**. It represents the velocity magnitude and pressure contours in the top row and the velocity streamlines and glyphs in the second row. As discussed in the previous section that the tip of the conical spike is at the maximum velocity and tends to decrease with time. The surface tension forces dominate the structure of the conical spike and plays a vital role in the deceleration of the spike tip. The momentum diffusivity at the tip and competing surface tension can be critically analyzed by visco-capillary scaling where the scaling regime corresponds to Capillary number ($Ca = \mu U/\sigma$), where $U$ is the characteristic velocity that is considered to be the initial velocity of spike, i.e, 4.04 m/s at t = 0.0225 s. In this case, the $Ca$ is evaluated to be 0.056 (in the $O(10^{-2})$). It further reveals that the surface tension forces dominate the mechanism and further validates the results discussed in the previous section (i.e., **sec. 4.1**). The surface tension forces thus cause the high momentum liquid in the tip to form the bulbous structure leading to the necking mechanism as shown in **Fig. 8(a-d)** at t = 0.0297 s. The necking phenomenon provides rich dynamics of the drop vibration and atomization with respect to the velocity, pressure, and vortex interactions as discussed in the next section (i.e., **sec. 4.2.1**).

### 4.2.1. Dynamics of necking event

**Fig. 9** elucidates a plot for the dynamics of the necking event till the primary pinch-off. At the initial stage of the necking phenomenon (i.e., at t = 0.02766 s), the tip pressure is higher than the neck pressure as shown in **Fig. 9**. However, the tip and neck velocity magnitude are observed to be almost equivalent as the high momentum liquid is forming the bulbous structure during the necking event. As time progresses, the neck becomes thinner, and the neck pressure increases. On the other hand, the tip pressure is observed to decrease as the surface tension forces compete the inertial and viscous forces. The velocity magnitude decreases in both cases, but the tip velocity lowers than the neck velocity magnitude due to a local high neck pressure zone formation (see **Fig. 8(b, c)**). From t = 0.0292 to 0.0302 s, the neck velocity magnitude flattens as the difference between the tip and neck pressure reduces as shown in **Fig. 9**. A visualization of the phenomenon can be observed in **Fig. 8(b)**, where the tip and neck pressure are almost equal and high velocity magnitude is at the neck. At t = 0.0302 s, a crossover in the pressure zone is observed as the neck pressure surpasses the tip pressure due to high liquid flow through a smaller cross-section area via the neck. The reversal in the local pressure zones due to the necking event induces an increasing velocity magnitude at the neck at t = 0.0312 s. Further, decrease in the neck diameter then increases the neck pressure drastically with respect to the tip pressure and the neck velocity increases from 0.6 to 0.7 m/s at the onset of pinch off as observed in **Fig. 9**. A qualitative visualization can also be drawn from **Fig. 8(c)**, where it shows that the velocity magnitude and pressure is undisputedly maximum at the neck. The direction of pressure gradient in z-axis is then changes at the neck zone. In this way, the inertial forces via pressure gradient play a role (but not dominant) in the pinch-off mechanism. The bulbous structure at the tip pinches off to move upward whereas the rest of the drop retracts downward as a ligament as shown in **Fig. 8(d)**.

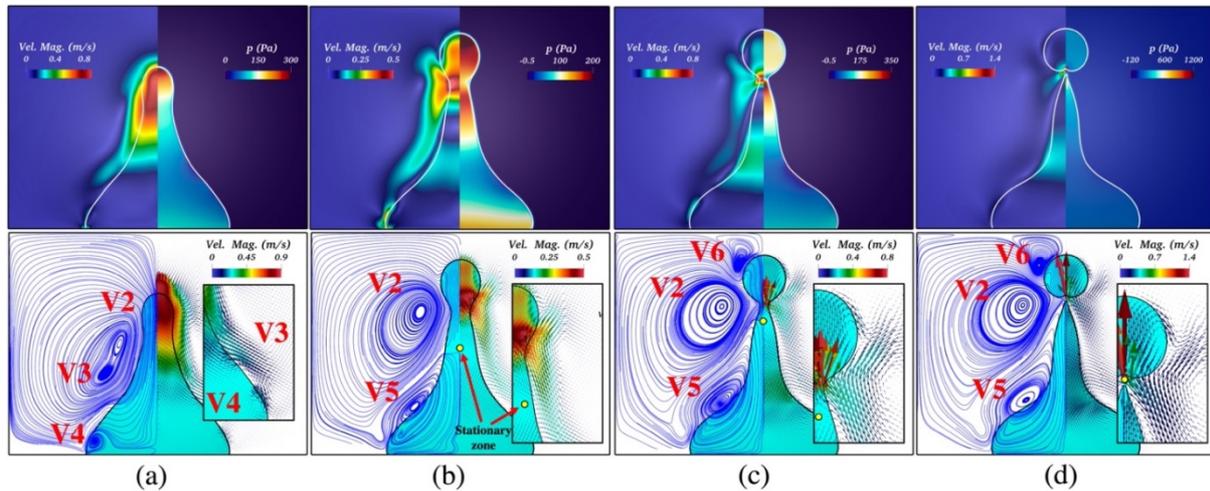

Fig. 8. Snapshots of the pinch-off mechanism from t = (a) 1.65T (0.0271 s), (b) 1.8125T (0.0297 s), (c) 1.906T (0.0312 s), and (d) 1.938T (0.0318 s) respectively: (top) Velocity magnitude and pressure contours (bottom) Velocity streamlines and glyphs.

The vortex interactions in the necking phenomenon are illustrated in the bottom row of **Fig. 8** by the velocity streamlines and glyphs. At the onset of the necking phenomenon (i.e., the initial state of the bulbous structure), a vortex disintegration is observed (V3 from V2) due to a new curvature formation via necking (see **Fig. 8(a)**). In the later stage (i.e., at t = 0.0297 s), the V3 vortex dissipates and the V4 vortex in opposite direction grows to form V5 as shown in streamlines and glyphs in **Fig. 8(b)**. The opposing vortices then create a stationary point (marked yellow in **Fig. 8(b)**. The stationary point creates a zero-velocity field whose neighborhood velocity field is in opposite direction. As time progresses, the neck diameter decreases and the negative curvature of the spike increases. Hence, the vortex V5 grows, and the stationary point moves in the upward direction as shown in **Fig. 8(c)**. At this interval, as the bulbous structure gets prominent a new vortex V6 is observed near the tip of the drop. Finally, at the time of pinch-off, the stationary point is observed to be at the pinching off position as shown in **Fig. 8(d)**. It is thus elucidating the velocity field in opposite direction as discussed earlier.

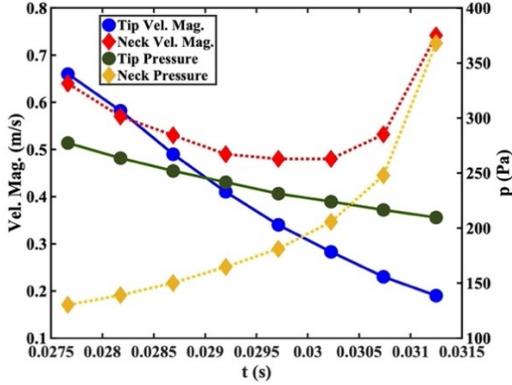

Fig. 9. Comparison between the tip and neck pressure and velocity magnitude during the necking stage at primary pinch-off

### 4.2.2. Universal primary pinch-off mechanism

The pinch-off test is conducted by computing the $r_{min}$ as the minimum neck radius as follow: two general points (S1 and S2) are considered on the axis of the drop along z-direction in such a way that the necking event occurs in between these two points. Next, the distance from the axis to the interface is obtained (i.e., $r$) and the minimum value is obtained as $r_{min}$. Finally, as we know that the neck radius should be $r_{min}$ from the interface profiles illustrated in **Fig. 10**, the neck radius is thus obtained.

The pinch-off profile is then obtained by defining a backward clocktime from the pinch-off interval (i.e., $t_p = 0.0312$ s). If $t$ is the forward clocktime, the difference between the $t_p$ and $t$ then provides the backward clocktime. The pinch-off dynamics profile obtains a slope equal to 1 following the universal scale law of visco-capillary regime given by a *spreading-like law* as [54],

$$r = \delta_{vc}\frac{\sigma}{\mu}(t_p - t) \quad (4.4)$$

Here, $\delta_{vc}$ corresponds to constant Capillary number regime. To further validate the arguments drawn from the pinch-off test profile, it is essential to analyze the interplay between the inertial, viscous and surface tension forces. Therefore, the choice of Ohnesorge number ($Oh$) regime is suitable for the investigation of the pinch-off test argument. $Oh$ provides information about the inertial and viscous forces in conjunction with the surface tension forces given by, $Oh = \mu/(\rho\sigma D)^{1/2}$. Here, $D$ is the extrinsic size (characteristic length) of the phenomenon. It is beneficial to understand that $Oh$ is related to the three important intrinsic parameters and one extrinsic parameter of the fluid. In the pinch-off mechanism, the fluid dynamics community usually chooses either the nozzle diameter (in case of jets), or the initial neck length as the extrinsic size $D$. The latter is implemented to obtain the extrinsic size in our case. At t = 0.0266 s, the necking event begins, and the size of the neck is obtained as $8.9 \times 10^{-4}$ m. The intrinsic properties of the water and the extrinsic size $D$ as the initial neck length, the $Oh$ is equal to $3.97 \times 10^{-3}$. It is to be noted that the lower value of $Oh$ should not be misguided as a inertio-capillary regime as the intrinsic factors influencing the inertial and surface tension forces are present in the denominator of $Oh$. Therefore, the different regime time scales are evaluated to further investigate the dominant forces. A fourth system of units can be derived from the intersection of visco-capillary ($vc$), inertio-capillary ($ic$) and inertio-viscous ($iv$) regime in the $Oh$ scale as,

$$\tau_o = \frac{\mu^3}{\sigma^2\rho} \quad (4.5)$$

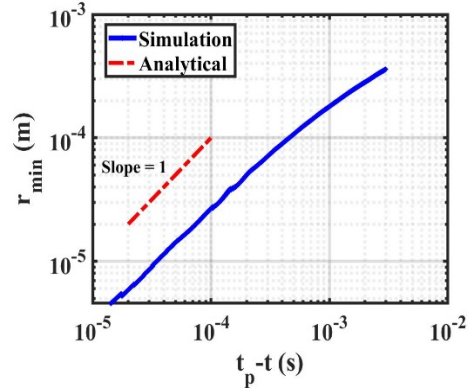

Fig. 10. Universal primary pinch-off test for the necking radius calculation

Where, $\tau_o$ is the intrinsic time scale which is called the $Oh$ time unit. Further, the time scales for the three regimes (i.e., $\tau_{vc}, \tau_{ic}, \tau_{iv}$) can be obtained by [54],

$$\tau_{iv} \xrightarrow{\times Oh} \tau_{ic} \xrightarrow{\times Oh} \tau_{vc} \xrightarrow{\times Oh^2} \tau_o \quad (4.6)$$

Using **Eq. 4.6**, $\tau_{vc}, \tau_{ic},$ and $\tau_{iv}$ are found in the $10^{-5}$, $10^{-3}$, and $10^{-1}$. Among the three time scales, it is evident that the pinch-off is governed by the visco-capillary regime as the pinch-off event occurs in the $10^{-5}$ scale as shown in **Fig. 10**. Therefore, it is concluded that the viscous and surface tension forces represent the universal pinch-off mechanism for the primary pinch-off event and the effects of the vibration of the surface via inertial body forces do not govern the dynamics of the pinch-off.

However, it is to be noted that the characteristics of the vibration (frequency and wavelength) govern the dynamics of detached drop motion. Once the pinch-off occurs, the bulbous structure shoots upward. In the frame of reference of the observer, the velocity at which the bulbous drop shoots ($v_{bulbous}$) is given by,

$$v_{bulbous} = v_{t_p} + \frac{A\sin(2\pi f t_p)}{2\pi f} \quad (4.7)$$

where, $v_{t_p}$ is the average velocity of the bulbous structure in the frame of reference of the vibrating drop. For simplicity, $v_{t_p}$ is assumed to be equal to the neck velocity magnitude at the onset of pinch-off mechanism (i.e., 0.72 m/s). Using **Eq. 4.7** at $t_p = 0.0312$ s, the $v_{bulbous}$ is equal to 0.621 m/s. Assuming the drop as a point mass governed by kinematics, the maximum height reached by the atomized drop is 2 cm (i.e., 8x of the vibrating sessile drop radius). After the bulbous structure atomizes, the rest of the spike subsequently retracts like a ligament and a new bulbous region is formed for pinch-off. This event is discussed in detail in the next section (**sec. 4.3**).

### 4.3. Ligament retraction and secondary pinch-off

The primary pinch-off is followed by the formation of a cylindrical ligament which retracts back to the vibrating drop. **Fig. 11(a)** represents the ligament retraction phenomenon under the influence of the vibration where the pressure and velocity magnitude, streamlines and glyphs as. The phenomenon is like the ligament retraction works done previously by Basaran et al. [55] and Constante-Amores et al. [56]. The latter works are done with the same code platform which is used in the present work. However, the difference between the previous works and the current work is the morphology of the ligament. They utilized a single cylindrical ligament retracting from north and south tip forming capillary waves whereas in the present work, the cylindrical ligament retracts back to the vibrating drop. Moreover, the present work is under time-varying oscillatory body forces whereas the previous works are not induced with such forces.

Having said that, a comparison between the dynamics of both the problems can draw interesting conclusions as follow: First, in an initially motionless cylindrical ligament, the north and south tip starts to retract due to the pressure gradient between these tips (high pressure zones) and the body of the ligament (low pressure zones). Henceforth, the inertial forces help in the initial movement of the ligament. Similarly, in the case of ligament retraction of a vibrating drop, the inertial forces are found to dominate for the retraction of the ligament in the form of pressure gradient and negative acceleration body forces. However, in this case, the high pressure zone is observed at the tip of the ligament whereas the low pressure zone is present at negative curvature connecting the vibrating drop to the ligament (see **Fig. 11(a)**). Second, in the previous works [55], [56], the two tips retract and showcase capillary waves via surface tension forces. It is analogous to the present case where the surface tension forces dictate the high momentum fluid into a bulbous structure as shown in **Fig. 11(b)** and a secondary neck is formed. This mechanism is similar to the necking of conical spike in the previous section. Third, by expanding the Taylor-Culick expression from an axisymmetric retracting liquid sheet to a cylindrical ligament (i.e., $V = \sqrt{2\sigma/\rho\pi R}$), Constante-Amores et al. [56] found excellent validation of the tip velocity. In our case, however, taking $R$ as the initial radius of the ligament, we obtain 0.3517 m/s in the frame of reference of the observer ($V' = V + A\sin(2\pi f t)/2\pi f$), which is 50% lesser than the initial velocity observed in the simulation, (i.e., 0.704 m/s see **Fig. 12** for tip velocity at initial stage). The difference can be understood by the additional inertial forces the ligament acquires by virtue of oscillatory body forces due to forced vibration.

One of the interesting characteristics of the cylindrical retracting ligament is the secondary necking event as shown in **Fig. 11 (b, c)**. The pressure gradient from the tip of ligament magnifies as it reaches the negative pressure zone (see -10 Pa in **Fig. 11(a)**). An increase in the velocity magnitude is also observed at the intersection

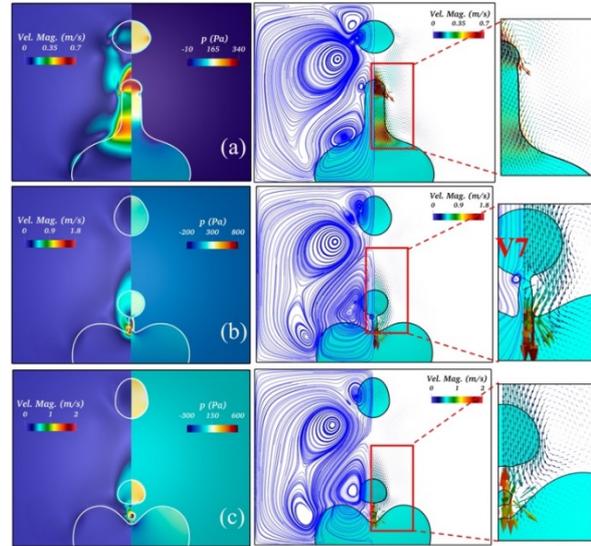

Fig. 11. Velocity magnitude and pressure contours (left) and velocity streamlines and glyphs (right) for (a) Ligament retraction after pinch-off mechanism at t = 2.031T (0.0332 s) (b) secondary pinch-off mechanism at t = 2.093T (0.0348 s) (c) satellite drop formations at 2.123T (0.0353 s)

of the ligament with the drop due to the sudden increase in the cross-sectional area from the neck to the drop as shown in the contours of **Fig. 11(a)** and glyphs in **Fig. 11(a)**. Henceforth, the necking velocity magnitude starts to increase (irrespective of the pressure gradient) as the necking size decrease due the surface tension forces for the formation of the bulbous structure (see **Fig. 11(b)**). It can be understood from **Fig. 12** where at the onset of the necking event (i.e., 0.0332 s), the neck and tip velocity magnitude are observed to be almost same and as time progresses, the necking radius decrease and induces higher inertial forces to push the fluid from the neck to the drop of large cross-sectional area.

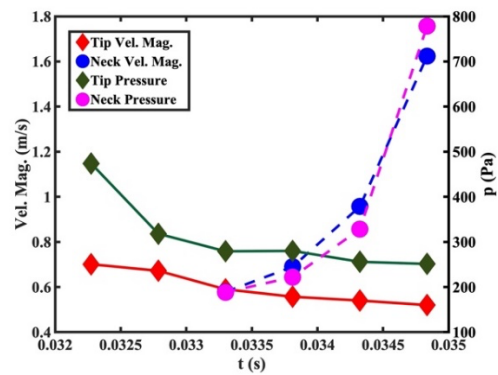

Fig. 12. Comparison between the tip and neck pressure and velocity magnitude during the necking stage at secondary pinch-off

Once the bulbous structure becomes prominent (see **Fig. 11(b)**), the neck pressure velocity magnitude is much higher than the tip and creates sufficient condition for the secondary pinch-off event. Meanwhile, a small vortex V7 (refer the inset of **Fig. 11(b)**) is formed near the top side of the neck which induces the pinch-off of the secondary drop from the neck. Moreover, it is observed that due to the high flux of liquid flow from the neck to the drop, the peripheral region become bulge and forms a secondary crater at the central region of the drop. It leads to a breakage of the neck into another smaller drop as a satellite drop as shown in the inset of **Fig. 11(c)**. This event takes place simultaneously and two drops are ejected which eventually falls back to the surface of the drop.

### 4.3.2. Universal secondary pinch-off mechanism

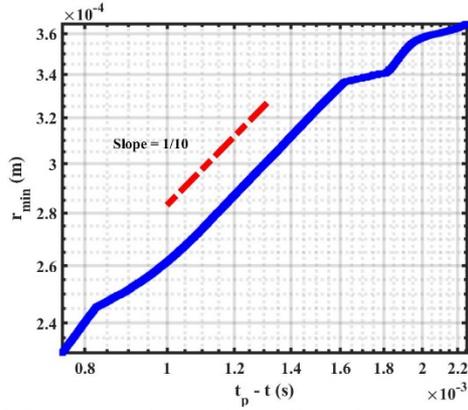

Fig. 13. Universal primary pinch-off test for the necking radius calculation

Similar to the pinch-off test in **sec. 4.2.2**, another pinch-off test is conducted for the secondary pinch-off (i.e., the pinch-off at the top side of the neck). In this case, the slope follows a different scaling i.e., 1/10, (see **Fig. 13**). The decrease in the slope and not following the visco-capillary regime is due to several reasons: (i) the inertial forces are found to play significant role in the liquid flow from high momentum tip of the ligament towards the vibrating drop via a radially decreasing neck size, (ii) no stationary point movements are observed as compared to the primary pinch-off which distinguishes the axial flow in opposite directions, (iii) the pinch-off occurs as a small opposite vortex V7 due to the curvature is formed near the tip of the neck and the secondary bulbous structure, and mainly (iv) the inertial forces via oscillatory body forces can deviate the pinch-off regime significantly in this case (v) similar $Oh$ analysis with $D$ as initial ligament diameter showed that the time scale is $10^{-3}$ s- suggests role of inertial forces.

The above reasons showcase that the universal pinch-off scaling does not correspond to the visco-capillary regime and the inertial forces become important in the secondary pinch-off mechanism. It is evident from the above discussion that the inertial forces drive the pinch-off and once the secondary bulbous structure is ejected, it has a downward velocity as shown in the glyphs of **Fig. 11(c)**. It means that these ejected drops rather than shooting as the primary election, falls back and coalesce with the vibrating sessile drop. Hence, the primary and secondary pinch-off mechanism not only differs from their attributes of post-ejection, but also differs in their pinch-off regime.

## 5. CONCLUSION

In this work, we elucidated two major open questions in vibrations of sessile drops using DNS- first, what causes subharmonic azimuthal waves on a vibratory sessile drop?

and second, what universal regime the pinch-off event occurs under forced vibrations? To explain the first question, we considered high mode excitation with a large sessile drop of 100 $\mu L$. It is elucidated that rather than the harmonics of the solid vibrations, it is the harmonics of the interfacial waves itself that produces subharmonic response near the contact line. It is further illustrated as a tip and contact line dynamical system where the mode of dynamics changes when the sessile drops undergo from axisymmetric to coupled axisymmetric-azimuthal waves. Such a dynamical system is analogous to parametric driven double pendulum model where the harmonics and subharmonics coincide. In order to answer the second question, a low mode excitation of a smaller drop of 30 $\mu L$ is considered. It illustrated as proof-of-concepts to the experiments which conclude that the crater structure is required for the drop atomization. Five major events are critically analyzed, i.e., crater, spike formations, primary pinch-off, ligament retraction and secondary pinch-off. The analysis of the five events provided sufficient foundation to explain the universal regime of the pinch-off mechanism. In case of primary pinch-off, a visco-capillary regime is observed and explained with the help of $Oh$. However, the secondary pinch-off showcased a different regime and can not be attributed to a universal regime due to the effects of forced oscillations on the ligament. It forms a secondary and a satellite drop which eventually falls back to the vibrating drop.

The future works for high mode excitation involves the atomization of the vibrating sessile drop where the azimuthal waves begin to grow over the axisymmetric waves to produce chaotic wave modes. The troughs of coupled axisymmetric and azimuthal eventually becomes craters similar to the case of low mode excitation and thus, follows similar atomization as discussed in this work. There are two ways to achieve such simulation- First, introduce a random seeding to the waves formed in high mode excitation, where the maximum uncertainty is present at the tip and minimum near the contact line. Seeding can eventually lead to the surface breakup, but the chaotic mixing is not formed physically by letting the interfacial waves to go through all wave modes as done to obtain azimuthal waves in this work. On the other hand, the second way is to linearly increase the acceleration amplitude from 100 m/s$^2$ to 2000 m/s$^2$ over a time interval of 20T and then keep the acceleration constant at 2000 m/s$^2$. In this way, the drop undergoes all the wave modes required to produce chaotic behavior due to the superposition of axisymmetric and azimuthal waves. The trade-off in this step is the mesh setup. As the wavelength is in the order of $10^{-5}$ and the amplitude acceleration is in the order of $10^3$, a finer mesh is required and hence, more computational cost. Both approaches will be implemented, compared and assessed in a companion paper in the future.

## 6. ACKNOWLEDGEMENTS

I would like to thank my family and friends for their support throughout my academic year. I convey special thanks to Dr. Lyes Kahouadji for introducing me to the beauty of Faraday waves and fluid instabilities. I also extend my regards to Prof. Laurette Tuckermann, whose insights to the Faraday waves enabled me to distinguish my master thesis to a new direction. Finally, I am thankful to Imperial College London Computing Facilities to provide allocated computing time.

## 7. SUPPLEMENTAL MATERIAL

The figures illustrated in this work can be accessed in high resolution at
https://www.dropbox.com/home/MSc%20Thesis/Thesis%20Figures.

## 8. APPENDIX

### A.1. Velocity over the harmonic interfacial wave

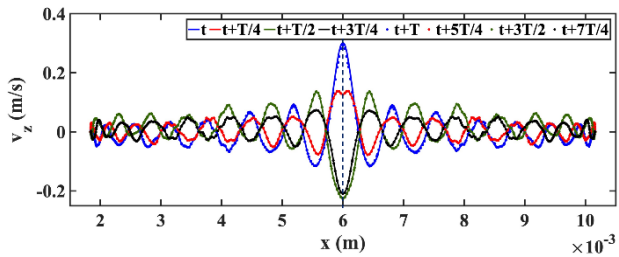

Fig. A.1.1. $z$-velocity field over the interface for axisymmetric waves

Similar to the argument to the pressure fields discussed in **sec. 3.1**, the $z$-velocity field is also observed to be axisymmetric with respect to the dotted line (i.e., the axis), harmonic as the velocity field is repeating after every interval T, and standing as the maximum amplitude is same after every time T.

### A.2. Qualitative validation via visualization

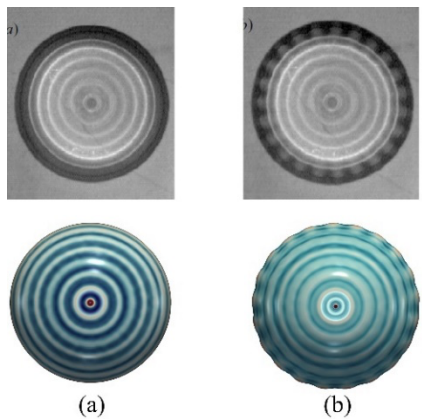

Fig. A.2.1. Visualization of the high mode excitation for (a) axisymmetric and (b) axisymmetric-azimuthal by (top) experiments [24] (bottom) our simulation

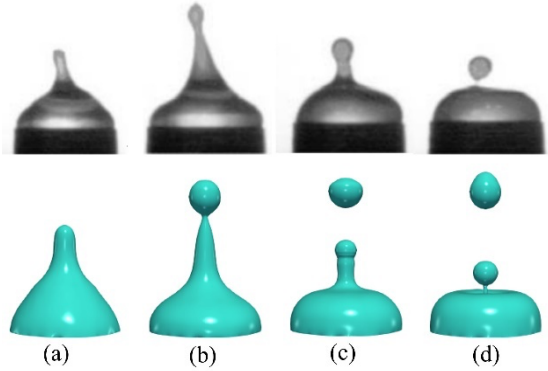

Fig. A.2.2. Visualization of the low mode excitation for (a) conical spike, (b) primary pinch-off, (c) ligament retraction, and (d) secondary pinch-off by (top) experiments [49] (bottom) our simulation

*Corresponding Supervisor: *o.matar@imperial.ac.uk*